\documentclass[12pt,preprint]{aastex}

\shorttitle{Imaging of Polarized SiO High $J$ Maser towards VY CMa}

\shortauthors{Shinnaga et al.}

\received{2004 May 12}
\begin{document}


\title{Interferometric Observation of the Highly Polarized   
     SiO Maser Emission from the $v=1, J=5-4$ Transition  
     Associated with VY Canis Majoris}



\author{Hiroko Shinnaga\altaffilmark{1,2}, 
James M. Moran\altaffilmark{3}, 
Ken H. Young\altaffilmark{3}, and 
Paul T.P. Ho\altaffilmark{3}}
\email{hshinnaga@cfa.harvard.edu}
\altaffiltext{1}{Harvard-Smithsonian Center for Astrophysics, Submillimeter Array, 
645 North A\arcmin ohoku Place, Hilo, HI 96720 USA
}
\altaffiltext{2}{Present address: California Institute of Technology  
Submillimeter Observatory, 111 Nowelo Street, Hilo, HI 96720 USA
}
\altaffiltext{2}{Harvard-Smithsonian Center for Astrophysics,
60 Garden Street, Cambridge, MA 02138, USA
}

\begin{abstract}
We used the Submillimeter Array  to image the SiO maser emission in the $v=1$, 
$J=5-4$ transition associated with the peculiar red supergiant VY Canis 
Majoris. We identified seven maser components 
and measured their relative positions and linear polarization properties.
Five of the maser components are coincident to within about 150 mas
($\sim$ 200 AU at the distance of 1.5 kpc); most of them may originate in 
the circumstellar envelope
at a radius of about 50 mas from the star along with the SiO
masers in the lowest rotational transitions. Our measurements
show that two of the maser components may be offset
from the inner stellar envelope (at the 3$\sigma$ level of significance) 
and may be part of a larger
bipolar outflow associated with VY CMa identified by Shinnaga et al.
The strongest maser feature at a velocity of 35.9 kms$^{-1}$ has a 60 percent
linear polarization, and its polarization direction is 
aligned with the bipolar axis. Such a high degree of polarization
suggests that maser inversion is due to radiative pumping.  Five of the other 
maser features have significant linear polarization.
  
%

\end{abstract}




\section{INTRODUCTION}
The unusual star VY Canis Majoris (VY CMa), 
is located 
at the edge of the very large ($\sim$ 5$^{\circ}$ in diameter) 
Sharpless 310 HII region 
at a distance of 1.5 kpc (e.g. \citet{her70,lad78}), and 
is one of the most intrinsically luminous stars known 
in the Galaxy with $L_* \sim (2 - 5) \times 10^5 L_{\odot}$ 
(e.g. \citet{les96}).  
Despite its high luminosity, it has a low effective temperature of $T_* \sim 2800 
{\rm \ K}$, 
making it a red supergiant of spectral 
class M5Ib with a mass of $\sim$ 25 M$_{\odot}$.  
Its unusually high mass loss rate  
for a red supergiant, $\dot{M}_* \sim (1 - 3) \times 
10^{-4} M_{\odot} {\rm \ yr^{-1}}$ \citep{jur90}, 
 is consistent with its thick circumstellar envelope 
and suggests that VY~CMa may be close to the end of its life cycle.  
High resolution images taken at optical and infrared wavelengths
reveal complicated circumstellar dust structures 
\citep{mon99,smi01}.
However, it is difficult to investigate the innermost circumstellar structures 
via the continuum emission 
because of the high opacity. 

VY CMa's extended outflowing envelope is traced by
various species of molecular masers that conform to 
the standard paradigm for late type and supergiant stars
whereby various 
species trace specific layers of the envelope for reasons involving
density and excitation process 
(e.g., \citet{mor77,rei78,mar98,shi03,miy03}).
One of the unusual characteristics of the emission from
the circumstellar envelope is the high degree of linear polarization
in the SiO $v=0$ lines. 
The only other stars that are known to show polarization in the
SiO ground state are R Cas \citep{shi02} and Orion KL/IRC2 source I 
\citep{tsu96}. 

We report the first 
imaging 
for the $v=1,~J=5-4$  transition with high angular resolution and 
also describe the spectral characteristics of the
strong linear polarization detected in this line.

\section{OBSERVATIONS AND DATA REDUCTION}
Observations of the SiO emissions in the $v=1,~J=5-4$ transition 
(215.595950 GHz) 
towards VY CMa were made with the partially completed 
{\it Submillimeter Array}\footnote{The Submillimeter Array is a joint
project between the Smithsonian Astrophysical Observatory and the
Academia Sinica Institute of Astronomy and Astrophysics, and is
funded by the Smithsonian Institution and the Academia Sinica.} 
(SMA; \citet{ho04}),  
located 
on Mauna Kea,  
on 7 December 2002 under fair weather conditions 
(the zenith opacity $\tau_{\rm 225 GHz}  \sim$ 0.15 $-$ 0.2).   
The phase tracking center was set at 
RA~=~07$^{h}$ 22$^{m}$ 58.$^{s}$27, Decl.~=~--25$^{\circ}$ 46$\arcmin$ 03.$\arcsec$4 
(J2000) and the field of view (FOV) was 57$\arcsec$.    
A compact configuration (baselines ranging from 6.3 to 18.1 k$\lambda$) 
with three antennas was employed for imaging the maser line.  
The beam size was 13.9$\arcsec$ $\times$ 5.6$\arcsec$. 
The spectral window was centered at 
10.0 kms$^{-1}$ in $V_{\rm LSR}$.  
The frequency resolution was $\sim$ 400 kHz.   
The typical double sideband system temperature was 500 K.
We observed VY CMa along  
with a relatively bright nearby quasar, 0727--115 
($\sim$ 14$\arcdeg$ away), for the relative amplitude and astrometeic phase calibration,  
and Callisto for the absolute flux calibration.  
Precise phase calibration was achieved through the use of the maser component
at 35.9 kms$^{-1}$ as a phase reference.
The passband calibration was done on Jupiter.  
Visibility data were calculated with the MIR IDL package \citep{sco93}.  
The quasar amplitude  
showed no systematic variation with random fluctuation  
less than $\sim$ 10\% of the mean value, which   
reflects the accuracy of the amplitude calibration.  

Linear polarization measurements were accomplished 
by taking advantage of the diurnal rotation of the 
sky with respect to the axis of polarization
sensitivity of the Array. The SMA antennas have
alt-azimuth mounts, and the receivers, mounted at the 
Nasmyth focus following reflections from five intermediate 
mirrors, have fixed linearly polarized feeds. 
The final reflection before the receiver is achieved 
with a wire grid, which 
ensures that the overall polarization purity is 
better than 20 db. 
Because of the Nasmyth optics, the plane of the
receiver polarization outside the antenna with respect to
the local vertical is $45 - EL$, where $EL$ is the elevation angle
of the source. 
The feed position angle on the celestial
frame is (e.g., \citet{sma62})
\begin{equation}
PA {\rm (deg)}  = 45 - EL + {\rm sin}{\frac{{\rm cos}(\phi_{\rm LAT})\cdot{\rm sin}
(HA)}{{\rm cos}(EL)}}
\end{equation}
where $\phi$ is the latitude of the Array ($\sim$ 19$\arcdeg$49$\arcmin$27$\arcsec$) 
and $HA$ is
the source hour angle. 
In our experiment, $HA$ ranged from
$-3$ hours to 2.5 hours and $PA$ ranged
from $-25$ to 60\arcdeg, 
which was sufficient to determine the 
linear polarization characteristics of the maser emission.

\section{ANALYSIS, RESULTS, AND DISCUSSION} \label{resultdiscuss}
\subsection{Line profile and linear polarization of the maser}
\begin{figure}
\plotone{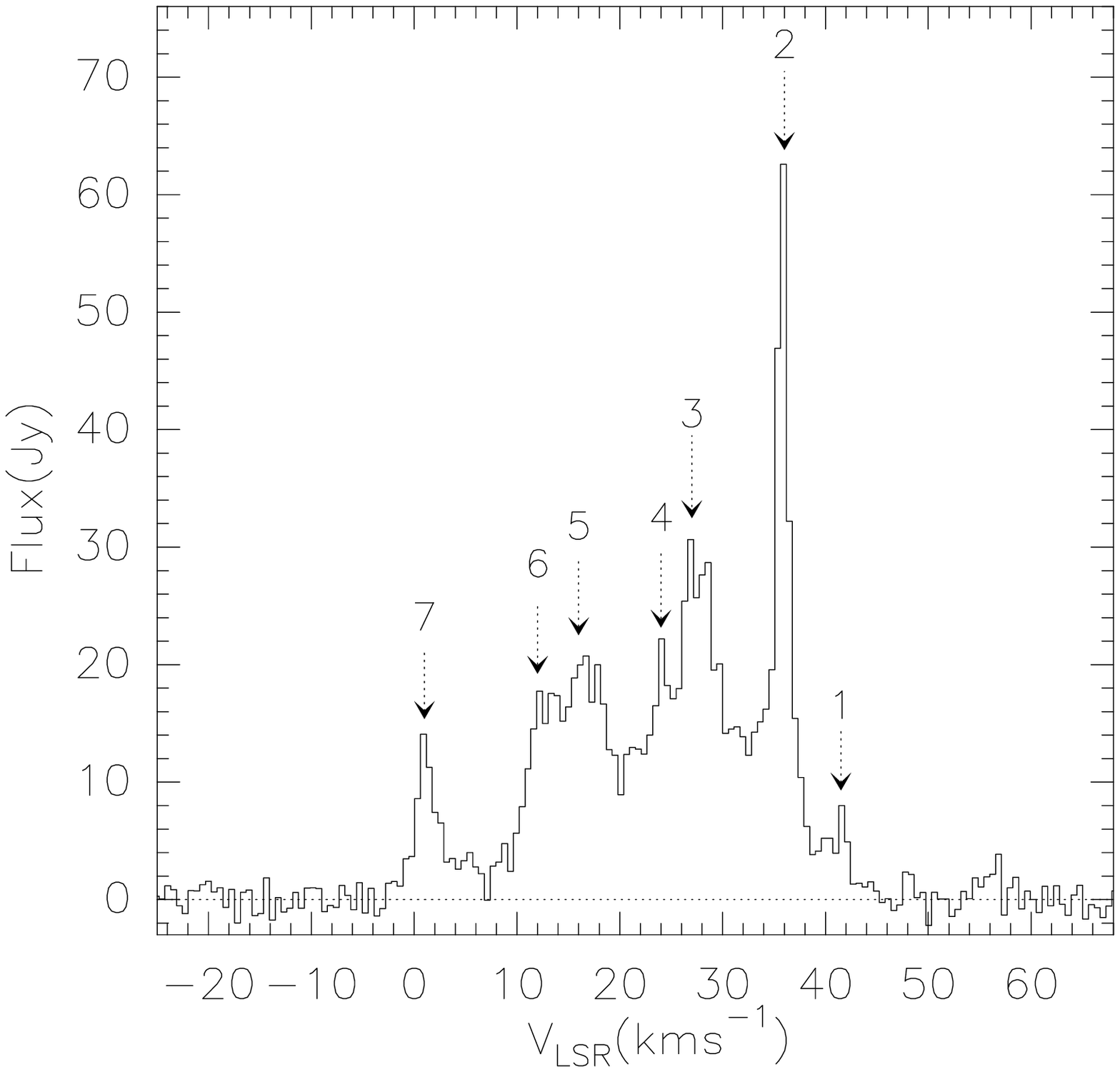}
\caption{An integrated spectrum of  
          the SiO high $J$ maser ($v=1$, ~$J=5-4$) 
          associated with VY CMa made from the sum of all
	  observations at various position angles. 
          The velocity axis (radio definition) is measured
          with respect to the Local Standard of Rest. The stellar velocity is
          about 20 kms$^{-1}$. There are probably additional components
	  at 48 and 56 kms$^{-1}$, which were too weak for 
	  detailed analysis.
          The numbers with arrows identify 
          the seven major components.  
          \label{specall}}
\label{fig:specall}
\end{figure}

Figure~\ref{fig:specall} shows an integrated spectrum (vector average of all
spectra) of 
the SiO high $J$ maser ($v=1$, $J=5-4$). 
The noise level in the spectrum was 1.1 Jy.     
Seven major velocity components have been identified by the number 
in the figure.   
In addition to these, we also detected 
two weak high velocity components at the red shifted edge, 
at $V_{\rm LSR}$ $\sim$ 48 and $\sim$ 56 kms$^{-1}$.  
The brightest maser, component 2, has a very narrow 
line width 
($<$  0.5 kms$^{-1}$ at full width at half maximum) corresponding to a 
thermal Doppler width of about 600K if the maser is saturated.  
The line profile of the high $J$ transition is quite different from 
those of the lower $J$ transitions (e.g., \citet{shi99}), 
indicating different regions for their origins.   
\placefigure{fig:specall}

\begin{figure}
\plotone{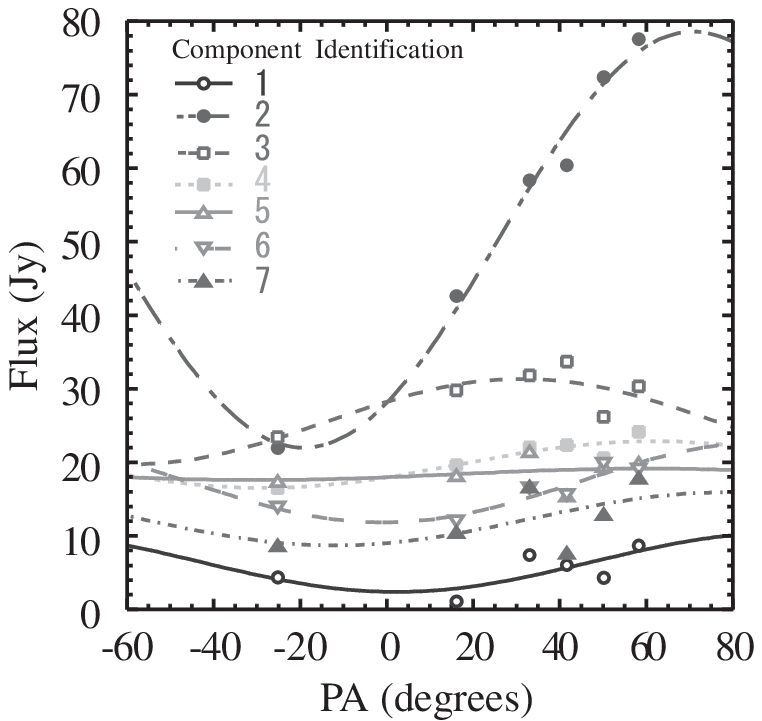}
\caption{The apparent flux density 
           of seven velocity components as a function of the position angle
	   of the feed polarization.  
           The flux density of 
           component 5 was almost constant over the $PA$s, 
           the flux of the other six components 
           changes as a function of $PA$ because of their linear polarization.   
           No fine gain adjustments were made in the data
	   presented here (see Section 3.1). \label{polariall}}
\label{fig:polariall}
\end{figure}

Figure~\ref{fig:polariall} the flux densities 
of the seven major components at each $PA$. 
To determine the polarization of the masers, we fit
the flux density of each maser component, $S_{k}$, as a function of  $PA$ to 
the equation
\begin{equation}
S_k = S_{0_k} + S_{p_k} \cos [2(PA-{\psi_k)}]
\end{equation}
where $S_{0_k}$ is the average flux density of maser component $k$, $S_{p_k}$ is the polarized
flux density, and $\psi_k$ is the position angle of the linear polarization.
Table 1 shows the results of this analysis. The fractional
linear polarization, listed in the rightmost column of Table 1,
is given by $S_{p_k}$/$S_{0_k}$. The fractional polarization percentages 
ranged from less than 10 percent (component 5) to about 60 percent
(components  1 and 2). These values of fractional polarization are lower limits
to the total polarization, since our measurement technique was not
sensitive to circular polarization.

Since component 5 showed no measurable change in
flux density over the 5.5 hour duration of the observations, we are
confident that the gain of the SMA was correctly calibrated throughout
the experiment by the normal calibration on the nearby quasar, and that this 
component has no measurable linear polarization.
To check this conclusion, and possibly improve the gain calibration, we
performed the following analysis. After the linear analysis (based on Eq. (2))
to determine
the three polarization parameters for each maser component, we examined
the residuals at each time ($PA$). We estimated a gain factor correction
for  each time (six parameters) to minimize the overall deviations. We
then solved again for the polarization parameters, and iterated the process
several times. The solution converged
quickly and stably, yielding the six gain factors: 1.00, 1.00, 1.04, 0.94,
0.99, and 1.03. The final polarization parameters were not significantly
different from those obtained in the first iteration (i.e., gain factors
set to unity). The iterative solution reduced the overall rms noise level
by about 20 percent and reduced the $\chi^2$ to essentially unity.
This demonstrates that the measurement of the polarization
parameters are accurate to within the quoted errors and are not
subject to systematic errors due to calibration errors in the
gain of the Array. It also shows that the $a$ priori gain calibration
based on the interleaved quasar observations was quite good, and
validates the implicit assumption that the maser components did
not vary intrinsically by more than a few percent over the period
of observations.  

We tested the hypothesis that the polarization angles of 
the maser spots are the same.  Under this assumption, the weighted mean 
polarization angle is 72 $\pm$ 4$\arcdeg$.  
The reduced $\chi ^{2}$ is 1.13, which strongly suggests that  
the polarization angles are the same and that the deviations from the mean 
of the position angles are not significant, except possibly for components 
3 and 6, which each deviate by about 1.5 $\sigma$.  
\placefigure{fig:polariall} 

\subsection{Angular distribution of the maser spots}
We assume that the individual masers components are unresolved 
by the SMA 
because their intrinsic sizes are less than ten milliarcsecond 
(e.g., see \citet{rei86} for discussion of maser angular sizes). 
We tried to measure the angular distribution of maser spots
with respect to the strongest maser component 2.
To estimate the positions, we made an image for each spectral channel 
and fit it with a two dimensional Gaussian function,
using AIPS tasks IMFIT and JMFIT.  To improve the component position error
estimates, we averaged the results over several channels
that contributed to the maser emission of each spot.
Note that, for a point source, the determination of the peak of 
the image is statistically equivalent to a $\chi^2$ fitting of the phases 
(e.g. \citet{tho01}).  
In particular, since the analysis
is based on the assumption that the individual maser components are point sources, 
the fact that the synthesized beam had high sidelobes was not important. 
Note that the visibility amplitude variations 
due to polarization do not bias the position estimate. 

Since we referenced the phase of each maser to the phase of the
reference feature, the effects of any systematic errors were
eliminated to first order. Any random phase errors between
channels were reflected in the formal position errors. However, any systematic
phase drifts between channels, caused by delay changes or
electronic changes in the filter responses could have affected the
measured positions. We have analyzed the phase versus time data
for the reference quasar, spectrally resolved into four channels, and
we found no evidence for any relative phase changes at the limit
of about two degrees, which was limited by the measurement noise.
The electronic stability and delay calibration of the system suggest that
the relative phase changes are much smaller than this amount.
A two degree systematic error would produce an error of less than 0.02$''$ 
in relative position.

Table 1 lists the positions of the masers, the relevant velocity ranges, and 
the measurement errors.
The errors are close to the expected levels predicted by the
formula $\delta \theta = 0.5 {\rm BW/snr}$ (e.g. \citet{kog96}),
where BW is the beamwidth in direction of $RA$ or Decl., and snr is the
signal-to-noise ratio. The snr of each component is its flux density 
divided by the measured noise of 1.10 Jy obtained from the integrated spectrum. 

Masers 3, 4, 5, and 7 are coincident with
maser 2 within the 1-$\sigma$ error bars (150 mas in each coordinate
or 200 AU). These masers are
most likely associated with the SiO masers in the circumstellar
shell at radius of 50 AU.  
Higher resolution measurements are
needed to determine their exact positions in the shell.
Masers 1 and 6 are offset in declination by 1300 mas (2.5$\sigma$) and 500 mas
(4$\sigma$), respectively. These offsets
might be real, but clearly have only marginal significance.
Although there appear to be no $J=1-0$ transition masers 
beyond the 50 mas shell \citep {miy03},
it is conceivable that masers could exist in the high $J$ transitions
in the bipolar outflow due to shock  
excitation. Masers 1 and 6 could be manifestations of such shocks.

The positions of the maser spots are plotted in Figure 3, with
the polarization indicated on each spot. 
We have placed component 2  
between the two outflow lobes at the position set by the central star.  
If the assumption is correct, the uncertainty of
the relative alignment of the two images is $\lesssim$ 0.1\arcsec. 
The relative alignment of
the $v=0$ image and our maser image is uncertain to $\lesssim$ 0.5\arcsec, 
based on the positions of the reference quasars observed in these two measurements.  
Note that
the position angle of the polarization of maser component 2 (71 $\pm$ 5$\arcdeg$) 
and that of the mean of all the components (72 $\pm$ 4$\arcdeg$)
are nearly aligned with the outflow.
This correlation suggests a relationship between the  
circumstellar envelope and the more extended bipolar outflow.
Note that since the components are on both sides of the systemic velocity,
they arise in both the approaching and receding sides of the envelopes.
%

\begin{figure}
\plotone{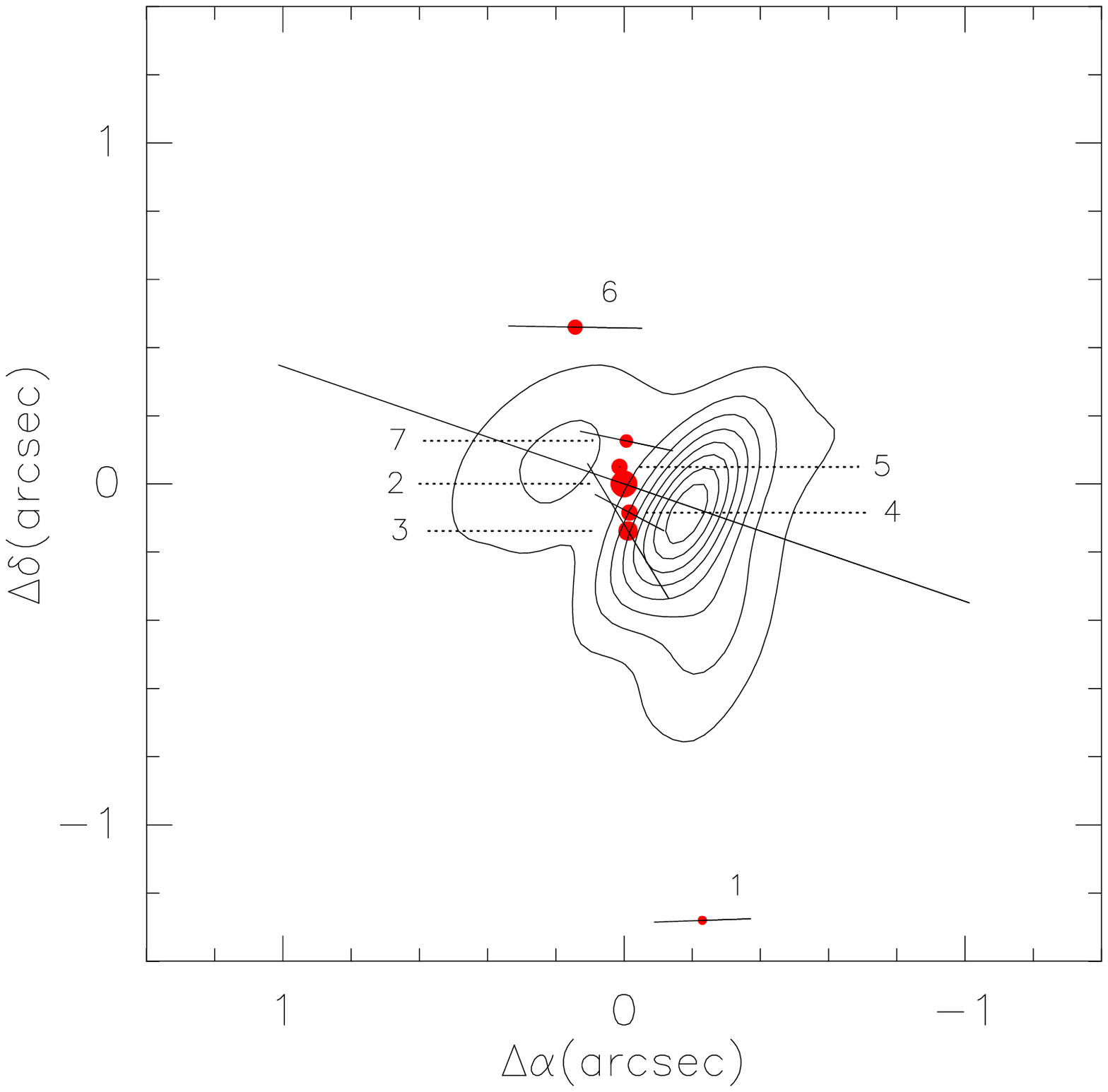}
\caption{Distribution of 
the SiO ($v=1,~J=5-4$) maser components, overlaid on the VLA image of the 
SiO ($v=0, ~J=1-0$) emission \citep{shi03}.  
The absolute position of maser 2 is RA(J2000) =  
7$^{h}$ 22$^{m}$ 58.31$^{s}$ $\pm$ 0.03$^{s}$, 
Decl.(J2000) = $-$25\arcdeg 46\arcmin 3.\arcsec2 $\pm$ 0.\arcsec5.  
The area of each spot is proportional to its flux density.  
The line lengths are proportional to the polarized flux density.  
Only maser 6 (the most northerly component),  
and possibly maser 1 (the most southerly component)
have statistically significant offsets from the position of maser 2 (see Table 1).
The mean polarization angle is 72 $\pm$ 4$\arcdeg$.  None of the deviations from 
this mean are significant, except perhaps for masers 3 and 6 (see Table 1).  }
\label{fig:polarimap}
\end{figure}

\placefigure{fig:polarimap}

\subsection{Continuum measurement}\label{comparison}
We imaged the continuum emission in the line free channels. Because
of the poor beamshape of the Array, we could not determine the spatial 
structure of the continuum. However, we measured the total
flux density to be 270 $\pm$ 40 mJy.  
The flux at 301 GHz and 658 GHz measured with the SMA recently were
340 $\pm$ 10 mJy and  
7.8 $\pm$ 2.6 Jy, respectively.   
The SED obtained by combining our results with other measurements in the submillimeter
band \citep{mar92,sop85} give a spectral index for flux density of about $-$ 3.0 for the
frequency range 200 $-$ 750 GHz.  
The emission is 
clearly almost entirely
due to dust, with a small contribution 
from the star. 
The spectral index is slightly less ($-$ 2.5) over the 
range 1,000 $-$ 15,000 GHz (300 $-$ 20 microns) \citep{les96}. 

\section{Pumping Mechanism of the SiO High $J$ Maser}\label{pumping}
Two theories have been proposed  to explain the SiO maser pumping
mechanism, especially for the low $J$ transitions: 
radiative pumping 
\citep{kwa74,deg76,buj94} 
and collisional pumping \citep{eli80,doe95}.    
The positional coincidence of emission among different $v \geq$ 1 
vibrational states as observed with KNIFE 
\citep{miy94} seems to support the collisional pumping theory.   

On the other hand, the 60$\%$ linear polarization that we observe
suggests radiative pumping \citep{ned90}. 
A fractional polarization this high for the $J=5-4$ transition
apparently requires anisotropy in the angular distribution of
the IR radiation involved in determining the populations of the
masing states \citep{wes84}.
This anisotropy could occur, in principle, for either collisional
(through the escaping IR radiation) or radiative pumping. However,
it occurs more naturally for radiative pumping since the incident
IR radiation comes from the direction of the star, and is a larger
effect in this case \citep{wes83}.  
Note that most theoretical analyses predict lower polarization
in the higher $J$ transitions, contrary to observations. 

The polarization direction of the radiation would be parallel or
perpendicular to the magnetic field if g$\Omega$ $>$ R, $\Gamma$, where
g$\Omega$ is the Zeeman rate, 1.5 rad s$^{-1}$ mG$^{-1}$, $\Gamma$ is
the maser level decay rate, about 5 s$^{-1}$, and R is stimulated
emission rate, 4 x 10$^{-6}$ $T_b$ $\Omega_m$, where $T_b$ is the
brightness temperature and $\Omega_m$ is the maser beam angle \citep{gol73}.  
Otherwise, the polarization would be
aligned with the direction of the pump anisotropy.  The VLBA image of
the SiO masers in the $v=1, J=1-0$ transition suggest that the spot
sizes are about 2 mas (about 2 AU) \citep{miy03}, 
so the maximum brightness temperature is about 4 x 10$^{8}$K. Taking
$\Omega_{m} = 10^{-2}$ (e.g., \citet{rei86}), 
the masers satisfy the g$\Omega \>$ R criterion if $B >$ 10 mG.
Since the magnetic field in the water masers in VY CMa
has been estimated to be about 150 mG, based on a Zeeman
interpretation of the observed circular polarization \citep{vle02},
polarization would be expected to trace the magnetic field. Note that
well ordered magnetic fields have been seen in other SiO masers
(e.g. VY CMa: \citet{shi03}, TX Cam: \citet{kem97}). 

\acknowledgments
We thank all SMA team members for their enthusiastic work 
and the Hawaiian people  
for allowing us to use their sacred mountain, Mauna Kea.    
We are grateful to 
Raymond Blundell, Shuji Deguchi, Ray Furuya, Liz Humphreys, Makoto Miyoshi, 
Scott Paine, and William Watson  
for helpful discussions.

\clearpage
                                                                                                                                  
\begin{deluxetable}{ccrcrrrc}
\tabletypesize{\scriptsize}
\tablecaption{Parameters of Seven Maser Components\label{tbl-1}}
\tablewidth{0pt}
\tablehead{
\colhead{Number} 
& \colhead{Velocity\tablenotemark{a}}
& \colhead{Velocity\tablenotemark{b}}   
& \colhead{Peak Intensity} 
& \colhead{X Offset$\tablenotemark{c}$} 
& \colhead{Y Offset$\tablenotemark{c}$}  
& \colhead{PA} 
& \colhead{Degree of Linear}  
\\
\colhead{} 
& \colhead{(kms$^{-1}$)}
& \colhead{Range}   
& \colhead{(Jy)}   
& \colhead{(arcsec)} 
& \colhead{(arcsec)}  
& \colhead{(degrees)} 
& \colhead{Polarization (\%)} 
}
\startdata
1 & 42.1   &41.6 - 42.7 &7.4   & $-$0.23 $\pm$ 0.30 & $-$1.28 $\pm$ 0.50   & 92 $\pm$ 30   &63 $\pm$ 31 \\
2 & 35.9   &33.1 - 37.6 &63.9  &0.00     &0.00      & 71 $\pm$ 5    &56 $\pm$ 4 \\
3 & 26.9   &25.2 - 30.3 &33.6  &$-$0.01 $\pm$ 0.06 &$-$0.14 $\pm$ 0.09     &31 $\pm$ 24  &23 $\pm$ 6 \\
4 & 24.1   &22.4 - 24.1 &23.7  &$-$0.02 $\pm$ 0.09  &$-$0.08 $\pm$ 0.13  & 62 $\pm$ 21   & 16 $\pm$ 4 \\
5 & 16.1   &13.9 - 17.3 &23.0  &0.01 $\pm$ 0.10 &0.05 $\pm$ 0.15   & ---  & $<$10 \\
6 & 12.2   &11.1 - 13.3 &20.2  &0.14 $\pm$ 0.07  &0.46 $\pm$ 0.10     & 89 $\pm$ 12   & 32 $\pm$ 6 \\
7 & 0.9    &$-$0.3 - 2.0&15.9  &$-$0.01$\pm$ 0.15 &0.13 $\pm$ 0.21   & 78 $\pm$ 60 & 29 $\pm$ 25 \\%
\enddata

\tablenotetext{a}{LSR velocity of maser component peak (used for polarization analysis).}
\tablenotetext{b}{Range of velocities used for position analysis.}
\tablenotetext{c}{The relative positions with respect to the brightest maser component, number 2.}

\end{deluxetable}                                                                                                               


\clearpage

\begin{thebibliography}{}
\bibitem[Bujarrabal(1994a)]{buj94} Bujarrabal, V.\ 1994, \aap, 
285, 971 
\bibitem[Deguchi \& Iguchi(1976)]{deg76} Deguchi, S.~\& 
Iguchi, T.\ 1976, \pasj, 28, 307  
\bibitem[Doel et al.(1995)]{doe95} Doel, R.~C., Gray, M.~D., 
Humphreys, E.~M.~L., Braithwaite, M.~F., \& Field, D.\ 1995, \aap, 302, 797  
\bibitem[Elitzur(1980)]{eli80} Elitzur, M.\ 1980, \apj, 240, 
553 
\bibitem[Goldreich, Keeley, \& Kwan(1973)]{gol73} Goldreich, 
P., Keeley, D.~A., \& Kwan, J.~Y.\ 1973, \apj, 179, 111 
\bibitem[Herbig(1970)]{her70} Herbig, G.~H.\ 1970, \apj, 162, 
557 
\bibitem[Ho, Moran, \& Lo(2004)]{ho04} Ho, P.~T.~P., Moran, J.~M.\&Lo, K.Y.\ 2003, \apj, 
in press(this issue)
\bibitem[Jura \& Kleinmann(1990)]{jur90} Jura, M.~\& 
Kleinmann, S.~G.\ 1990, \apjs, 73, 769 
\bibitem[Kemball \& Diamond(1997)]{kem97} Kemball, A.~J.~\& 
Diamond, P.~J.\ 1997, \apjl, 481, L111 
\bibitem[Kogan(1996)]{kog96} Kogan, L.\ 1996, NRAO AIPS Memo. Ser. 92 
\bibitem[Kwan \& Scoville(1974)]{kwa74} Kwan, J.~\& Scoville, 
N.\ 1974, \apjl, 194, L97 
\bibitem[Lada \& Reid(1978)]{lad78} Lada, C.~J.~\& Reid, 
M.~J.\ 1978, \apj, 219, 95 
\bibitem[Le Sidaner \& Le Bertre(1996)]{les96} Le Sidaner, 
P.~\& Le Bertre, T.\ 1996, \aap, 314, 896 
\bibitem[Marshall, Leahy, \& Kwok(1992)]{mar92} Marshall, C.~R., Leahy, D.~A.,
\& Kwok, S. \ 1992, \pasp, 104, 397 
\bibitem[Marvel, Diamond, \& Kemball(1998)]{mar98} Marvel, 
K.~B., Diamond, P.~J., \& Kemball, A.~J.\ 1998, ASP Conf.~Ser.~154: Cool 
Stars, Stellar Systems, and the Sun, 10, 1621 
\bibitem[Miyoshi et al.(1994)]{miy94} Miyoshi, M., Matsumoto, K., Kameno, S. et al.\ 1994, \nat, 371, 29 
\bibitem[Miyoshi(2003)]{miy03} Miyoshi, M.\ 2003, ASSL 
Vol.~283: Mass-Losing Pulsating Stars and their Circumstellar Matter, 303 
\bibitem[Monnier et al.(1999)]{mon99} Monnier, J.~D., 
Tuthill, P.~G., Lopez, B., Cruzalebes, P., Danchi, W.~C., \& Haniff, C.~A.\ 
1999, \apj, 512, 351 
\bibitem[Moran et al.(1977)]{mor77} Moran, J.~M., Ball, 
J.~A., Yen, J.~L., Schwartz, P.~R., Johnston, K.~J., \& Knowles, S.~H.\ 
1977, \apj, 211, 160 
\bibitem[Nedoluha \& Watson(1990)]{ned90} Nedoluha, G.~E.~\& 
Watson, W.~D.\ 1990, \apjl, 361, L53 
\bibitem[Reid \& Muhleman(1978)]{rei78} Reid, M.~J.~\& 
Muhleman, D.~O.\ 1978, \apj, 220, 229 
\bibitem[Reid \& Moran (1986)]{rei86} Reid, M., \& Moran, J. \ 1986, 
in Galactic and Extragalactic Radio Astronomy, 
eds. G.L. Verschuur \& K.I. Kellermann (Springer-Verlag), 255
\bibitem[Scoville et al. (1993)]{sco93} Scoville, N. Z., Carlstrom, J. E., 
Chandler, C. J., Phillips, J. A., Scott, S. L., Tilanus, R. P. J., 
\& Wang, Z. 1993, \pasp, 105, 1482
\bibitem[Shinnaga, Tsuboi, \& Kasuga(1999)]{shi99} Shinnaga, H., 
Tsuboi, M., \& Kasuga, T.  \ 1999, \pasj, 51, 175 
\bibitem[Shinnaga, Tsuboi, \& Kasuga (2002)]{shi02} Shinnaga, H., 
Tsuboi, M., \& Kasuga, T. \ 2002, IAU Symposium, 206, 278
\bibitem[Shinnaga et al.(2003)]{shi03} Shinnaga, H., 
Claussen, M.~J., Lim, J., Dinh-van-Trung, \& Tsuboi, M.\ 2003, ASSL 
Vol.~283: Mass-Losing Pulsating Stars and their Circumstellar Matter, 393 
\bibitem[Smart(1962)]{sma62} Smart, W.~M., Spherical Astronomy, 
Cambridge U. Press (Cambridge)
\bibitem[Smith et al.(2001)]{smi01} Smith, N., Humphreys, 
R.~M., Davidson, K., Gehrz, R.~D., Schuster, M.~T., \& Krautter, J.\ 2001, 
\aj, 121, 1111
\bibitem[Sopka et al.(1985)]{sop85} Sopka, R.~J., Hildebrand, 
R., Jaffe, D.~T., Gatley, I., Roellig, T., Werner, M., Jura, M., \& 
Zuckerman, B.\ 1985, \apj, 294, 242 
\bibitem[Thompson, Moran, \& Swenson(2001)]{tho01} Thompson, A., Moran, J., 
\& Swenson, G. Interferometry and Synthesis in Radio Astronomy, \ 2001 (Wiley, 
Second ed.)
\bibitem[Tsuboi et al.(1996)]{tsu96} Tsuboi, M., Ohta, E., 
Kasuga, T., Murata, Y., \& Handa, T.\ 1996, \apjl, 461, L107 
\bibitem[Western \& Watson(1983)]{wes83} Western, L.~R.~\& 
Watson, W.~D.\ 1983, \apj, 275, 195 
\bibitem[Western \& Watson(1984)]{wes84} Western, L.~R.~\& 
Watson, W.~D.\ 1984, \apj, 285, 158 
\bibitem[Vlemmings, Diamond, \& van Langevelde(2002)]{vle02}
Vlemmings, W.~H.~T., Diamond, P.~J., \& van Langevelde, H.~J.(2002) \ 2002, \aap, 394, 589
\end{thebibliography}
\end{document}